\documentclass[
twocolumn, 
reprint, 
superscriptaddress,
nofootinbib,
nobibnotes,
amsmath,
amssymb,
aps,
prb,
longbibliography,
]{revtex4-2}

\usepackage[justification=raggedright]{caption}
\usepackage{graphicx}
\usepackage{dcolumn}
\usepackage{bm}
\usepackage{xurl}
\usepackage{hyperref}
\hypersetup{colorlinks=true, allcolors=blue}   
\usepackage[capitalize]{cleveref}
\usepackage[T1]{fontenc}

\usepackage{amsmath}
\newcommand{\abs}[1]{\lvert #1 \rvert}

\usepackage{physics}   

\usepackage{xcolor}

\begin{document}

\preprint{APS/123-QED}

\title{Enhanced Coulomb pseudopotentials in high-pressure hydrides from first principles}

\author{Kieran Bozier}
    \email{kb697@cam.ac.uk}
\affiliation{%
 Department of Materials Science and Metallurgy, University of Cambridge, 27 Charles Babbage Road, Cambridge, CB3 0FS, United Kingdom
}%

 \author{Chris J. Pickard}
    \email{cjp20@cam.ac.uk}
\affiliation{%
 Department of Materials Science and Metallurgy, University of Cambridge, 27 Charles Babbage Road, Cambridge, CB3 0FS, United Kingdom
}%
\affiliation{
 Advanced Institute for Materials Research, Tohoku University, Sendai, 980-8577, Japan
}%

\date{\today}%

\begin{abstract}

In the Eliashberg theory of superconductivity, Coulomb interactions oppose the phonon-mediated attraction and suppress the superconducting critical temperature $T_{\mathrm{c}}$. However, most first-principles predictions continue to treat this repulsion with an empirical Coulomb pseudopotential $\mu^*_{\text{ME}}$ in the range 0.1--0.15. Here we present first-principles calculations of the Coulomb parameter $\mu$ and pseudopotential $\mu^*_{\text{ME}}$ in superconducting hydrides across a range of pressures. The screened Coulomb interaction is evaluated at the random phase approximation level with the \textsc{Mantel} code. We demonstrate that several hydride systems, including H$_3$S, YSbH$_6$ and Mg$_2$IrH$_6$, have strong Coulomb repulsions driven by a large density of states at the Fermi energy, $N_{\mathrm{F}}$. For Mg$_2$IrH$_6$ we obtain $\mu=0.62$ and $\mu^*_{\text{ME}}\approx0.39$, leading to a 40~K suppression in $T_{\mathrm{c}}$ relative to a standard $\mu^*_{\text{ME}}=0.16$. Our obtained $\mu^*_{\text{ME}}$ is far above the conventional range and, accounting for the energy dependence of the density of states, resolves the factor-of-two discrepancy between $T_{\mathrm{c}}$ estimates of previous studies. Across all systems and pressures studied, we find that $\mu$ closely tracks $N_{\mathrm{F}}$, suggesting that commonly used $\mu^*_{\text{ME}}$ values lead to overestimated $T_{\mathrm{c}}$ in materials where the density of states at the Fermi energy is large.


\end{abstract}

\maketitle

\section{Introduction}

Over the past decade, first-principles methods have played an increasing role in the discovery of novel superconducting materials. The search for conventional superconductors, which was historically driven solely by experiment, is now routinely guided by \textit{ab initio} calculations~\cite{boeri2021RoomtemperatureSuperconductivity2022, flores-livasPerspectiveConventionalHightemperature2020a}. This shift has been particularly important for high-pressure hydrides, where resolving their hydrogen positions experimentally remains a very challenging task~\cite{pickardSuperconductingHydridesPressure2020a}. Using a combined framework of electronic structure calculations, crystal structure prediction methods~\cite{pickard2006, glass2006, pickardInitioRandomStructure2011, lonieXtalOptOpensourceEvolutionary2011, wangCALYPSOMethodCrystal2012, oganovStructurePredictionDrives2019}, and Eliashberg theory~\cite{eliashberg_InteractionsElectronsLattice1960,scalapinoStrongCouplingSuperconductivity1966, marsiglioEliashbergTheoryShort2020}, several high critical temperature ($T_{\mathrm{c}})$ superconductors were first theoretically predicted and subsequently synthesized, including H$_3$S (203~K)~\cite{drozdovConventionalSuperconductivity2032015b, duanPressureinducedMetallizationDense2014, errea2015}, LaH$_{10}$ (250~K)~\cite{drozdovSuperconductivity250Lanthanum2019, kruglov2020, shipleyStabilitySuperconductivityLanthanum2020a}, and CaH$_{6}$ (214~K)~\cite{wangSuperconductiveSodalitelikeClathrate2012, liSuperconductivity200Discovered2022, houQuantumAnharmonicEffects2023}. Despite these successes, challenges remain to make the prediction of $T_{\mathrm{c}}$ truly \textit{ab initio}, as most estimates in the literature continue to rely on an empirical parameter, known as the Coulomb pseudopotential $\mu^*$.

Within Eliashberg theory, the superconducting properties are determined by an electronic self-energy with two contributions: an electron-phonon term and a screened Coulomb term. The first term, from the emission and reabsorption of a phonon, can be routinely evaluated using density functional perturbation theory (DFPT). This yields the Eliashberg spectral function $\alpha^2F(\omega)$ and electron-phonon coupling strength $\lambda$, and is the focus of most studies on conventional superconducting systems~\cite{margineAnisotropicMigdalEliashbergTheory2013}. The second term, involving the screened Coulomb interaction, is typically not treated with such care. In most work, the impact of this interaction is encapsulated in a Coulomb pseudopotential parameter $\mu^*$, which is usually taken as a constant between~0.1~and~0.15. This estimate derives from the Tolmachev-Morel-Anderson result~\cite{bogoljubov1958, morelCalculationSuperconductingState1962}, which assumes both the electronic density of states and screened Coulomb interaction are constant, and is justified primarily by the success of the Allen-Dynes formula in reproducing elemental $T_{\mathrm{c}}$ values. However, in principle the strength of this interaction should be materials-dependent, and the numerical value of $\mu^*$ further depends on whether the Allen-Dynes equation ($\mu^*_{\text{AD}}$) or the full Eliashberg equations ($\mu^*_{\text{ME}}$) are solved, with the latter introducing an additional dependence on the choice of Matsubara cutoff frequency \cite{pellegriniInitioMethodsSuperconductivity2024, koglerIsoMEStreamliningHighprecision2025a} and typically being much larger. As discussed in Ref.~\cite{pellegriniInitioMethodsSuperconductivity2024}, this distinction is often overlooked in practice, with Allen-Dynes pseudopotential values frequently used in the Eliashberg equations, which tends to result in an overestimate of $T_{\mathrm{c}}$.

Several works have sought to improve on the phenomenological $\mu^*$ values by employing first-principles methods. These typically calculate the Coulomb parameter $\mu$, which depends on screened Coulomb matrix elements and the density of states at the Fermi energy~$N_{\mathrm{F}}$, before renormalizing to obtain the pseudopotential~$\mu^*$. Early attempts by Lee and Chang confirmed that~$\mu^*_{\text{AD}}=0.1 $ was appropriate for Al~\cite{leeFirstprinciplesCalculationsCoulomb1995a}, while a larger~$\mu=0.49$ and ~$\mu^*_{\text{AD}}=0.14$ were estimated for Nb~\cite{leeLinearresponseCalculationCoulomb1996}, although the resulting $T_{\mathrm{c}}$ remained an overestimate. Moon et al. \cite{moon2004} used the same method for MgB$_2$ and observed that inhomogeneities in the charge density enhanced the dielectric screening, highlighting that $\mu$ is materials dependent. First-principles $\mu$ values were also calculated by Kawamura et al. for the elements~\cite{kawamuraBenchmarkDensityFunctional2020} and YNi$_2$B$_2$C~\cite{kawamura2017}. The most systematic efforts to date have been those of Sanna, Pellegrini and collaborators~\cite{flores-livasPerspectiveConventionalHightemperature2020a,pellegriniInitioCalculationsSuperconducting2023,sannaInitioEliashbergTheory2018,pellegriniInitioMethodsSuperconductivity2024,davydovInitioTheoryPlasmonic2020,sannaCombiningEliashbergTheory2020,pellegrini2019,pellegriniEliashbergTheoryAbinitio2022,sannaPredictionAmbientPressure2024,erreaQuantumCrystalStructure2020,mcfarlaneVanHoveSingularities2026,trescaWhyMercurySuperconductor2022}, who evaluate the static RPA screened Coulomb interaction using the all-electron \textsc{elk} code~\cite{elk, sannaInitioEliashbergTheory2018}. Across a range of systems, including layered superconductors such as CaC$_6$, MgB$_2$ and Li$_{0.5}$ZrNCl \cite{pellegriniEliashbergTheoryAbinitio2022}, they demonstrated that their first-principles $\mu$ and $\mu^*$ values yield more accurate $T_{\mathrm{c}}$ results than their phenomenological counterparts. In another work, they showed that although the RPA tends to underestimate the Coulomb repulsion in low-density systems compared to the Kukkonen-Overhauser (KO) approximation~\cite{pellegriniInitioCalculationsSuperconducting2023}, even these RPA-level results were a significant improvement on the phenomenological values. Despite these advances, the vast majority of studies on $\mu$ have focused on elemental metals and common conventional superconductors at low pressure.

First-principles estimates of $\mu$ and $\mu^*$ for hydrides, particularly at high pressure, remain scarce. Heil et al.~\cite{heil2019} obtained a value of  $\mu^*_{\text{ME}}=0.11$ for yttrium hydrides at 300~GPa, well in-line with phenomenological estimates. However, for other hydride systems this is not always the case, as demonstrated in Ref.~\cite{cerqueira2024}. In this work, which reported the most promising ambient pressure hydride systems from a screening of the \textsc{Alexandria}~\cite{schmidt2021, schmidt2023} database, they found the Coulomb parameter of the ten leading candidates varied from 0.58 for Mg$_2$IrH$_6$ to 0.16 for LiPdH$_2$. Although the resulting Coulomb pseudopotential $\mu^*_{\text{ME}}$ is lower than the Coulomb parameter $\mu$, and depends on factors such as the electronic bandwidth and Matsubara cutoff, this large range of values suggests a strong system dependence in the hydrides with values often exceeding the commonly used range.

The consequences of relying on standard $\mu^*$ values can be severe, as illustrated by the case of Mg$_2$IrH$_6$. This system is predicted to be a high-temperature conventional superconductor at ambient conditions, and was predicted independently by two groups using different methodologies. Ref.~\cite{doluiFeasibleRouteHighTemperature2024} employed carefully converged calculations using standard Eliashberg theory and a $\mu^*_{\text{ME}}=0.16$, obtaining a value of $T_{\mathrm{c}} = 160$~K. In contrast, Ref.~\cite{sannaPredictionAmbientPressure2024} employed superconducting density functional theory (SCDFT) to obtain a value of $T_{\mathrm{c}} = 71$~K. This technique includes the Coulomb effects from first-principles, although it requires additional approximations for the exchange-correlation functional and to extract other superconducting properties~\cite{kawamuraBenchmarkDensityFunctional2020, kruglov2020, pellegriniInitioMethodsSuperconductivity2024}. The discrepancy of a factor of two in the predicted $T_{\mathrm{c}}$ was suggested to arise either from disagreement in the calculated electron-phonon coupling strength, which is challenging in this material due to sharp features in the density of states, or from the treatment of the Coulomb interactions. While Ref.~\cite{doluiFeasibleRouteHighTemperature2024} used a standard $\mu^*=0.16$, first principles calculations in Ref.~\cite{sannaPredictionAmbientPressure2024} obtained a Coulomb parameter value $\mu=0.58$, and estimated this required a $\mu^*>0.25$.  

In this work, we investigate the Coulomb parameter $\mu$ across a range of hydride systems and pressures, including LaH$_{10}$, H$_3$S, CaH$_6$, YSbH$_6$, Mg$_2$IrH$_6$ and LaSc$_2$H$_{24}$. Calculations are performed with the recently developed \textsc{Mantel} code~\cite{mantel-v1}, which implements the methodology of Sanna and Pellegrini using \textsc{Quantum ESPRESSO}~\cite{giannozziQUANTUMESPRESSOModular2009} and \textsc{Yambo}~\cite{marini2009, sangalliManybodyPerturbationTheory2019}, and is validated against previous results for metallic elements and common superconductors. The impact of different exchange-correlation functionals is also considered. We explore the variation of $\mu$ across the hydride systems and as a function of pressure, and evaluate the resulting impact on the predicted $T_{\mathrm{c}}$ values.

\section{Theory}\label{sec:theory}

The expression for the static screened Coulomb interaction between two Bloch states $\ket{\mathbf{k}, n}$ and $\ket{\mathbf{k+q}, m}$ is given by~\cite{sannaInitioEliashbergTheory2018}
\begin{equation}
\begin{split}
    W_{\mathbf{k},n , \mathbf{k+q},m} &= \sum_{\mathbf{G} \mathbf{G'}} \bigg( \frac{1}{V}\frac{4\pi}{\abs{\mathbf{q+G}} \abs{\mathbf{q+G'}}} \epsilon^{-1}_{\mathbf{G},\mathbf{G'}}(\mathbf{q}, 0)\\
    &\quad \times 
    \rho_{\mathbf{k},n,\mathbf{k+q},m}(\mathbf{G}) \rho^*_{\mathbf{k},n,\mathbf{k+q},m}(\mathbf{G'}) \bigg)
\end{split}
\label{eqn:w_nkmp}
\end{equation}
where $\mathbf{k}$ is the initial wavevector, $\mathbf{k+q}$ is the final wavevector, $n$ is the initial band index and $m$ is the final band index. The summation is performed over reciprocal lattice vectors $\mathbf{G}, \mathbf{G'}$ and the scattering momentum $\mathbf{q}$ is restricted to the first Brillouin zone. The unit cell volume is denoted by $V$, and $\epsilon^{-1}_{\mathbf{G},\mathbf{G'}}(\mathbf{q}, 0)$ is the static inverse dielectric matrix, which is calculated here using the Random Phase Approximation (RPA). The terms $\rho_{\mathbf{k},n,\mathbf{k+q},m}$ are the plane wave matrix elements and are given by

\begin{equation}
    \rho_{\mathbf{k},n,\mathbf{k+q},m}(\mathbf{G}) = \bra{\mathbf{k+q}, m} e^{+i (\mathbf{q} + \mathbf{G})\cdot\mathbf{r}} \ket{\mathbf{k}, n}_{u.c.} 
    \label{eqn:rho}
\end{equation}
where the matrix element is evaluated as an integral over the unit cell. The symmetrized static inverse dielectric matrix is evaluated as
\begin{equation}
    \epsilon^{-1}_{\mathbf{G,G'}}(\mathbf{q},0) = \delta_\mathbf{G,G'} + \frac{4\pi}{|\mathbf{q}+\mathbf{G}||\mathbf{q}+\mathbf{G'}|} \;\chi^{red}_{\mathbf{G,G'}}(\mathbf{q},0)
    \label{eqn:inv_dielec}
\end{equation}
where the static reducible polarizability $\chi^{red}_{\mathbf{GG'}}(\mathbf{q},0)$  is defined as
\begin{equation}
\begin{split}
    \chi^{red}_{\mathbf{G,G'}}(\mathbf{q},0) = &\bigg[ \delta_{\mathbf{G,G''}} - \frac{4\pi}{(\mathbf{q}+ \mathbf{G''})^2}\, \chi^0_{\mathbf{G},\mathbf{G''}}(\mathbf{q},0) \bigg]^{-1} \, \\& \quad\quad\quad\quad \times\chi^0_{\mathbf{G''},\mathbf{G'}}(\mathbf{q},0) 
\end{split}
    \label{eqn:red_polar}
\end{equation}
and the non-interacting response function $\chi^0_{\mathbf{G},\mathbf{G'}}(\mathbf{q},0)$ within the RPA approximation can be written
\begin{equation}
\begin{split}
   & \chi^0_{\mathbf{G},\mathbf{G'}}(\mathbf{q},0)  = \\&    \     \frac{4}{N_{\mathbf{k}}V} \sum_{\mathbf{k}}\sum_{n}^{\mathrm{occ.}} \sum_{m}^{\mathrm{unocc.}} \frac{\rho_{\mathbf{k},n,\mathbf{k+q},m}(\mathbf{G}) \rho^*_{\mathbf{k},n,\mathbf{k+q},m}(\mathbf{G'}) } {\varepsilon_{\mathbf{k},n} - \varepsilon_{\mathbf{k+q}, m}},
    \end{split}
    \label{eqn:irred_polar}
\end{equation}
where $N_{\mathbf{k}}$ is the number of $\mathbf{k}$ points and $\varepsilon_{\mathbf{k},n}$ is the band energy of the $\ket{\mathbf{k},n}$ state, which here is taken as the DFT band energy.

The superconducting critical temperature is calculated using the isotropic Eliashberg equations. We use the variable density of states (vDOS) form of these equations as described in Ref.~\cite{koglerIsoMEStreamliningHighprecision2025a, lucreziFullbandwidthAnisotropicMigdalEliashberg2024a}, as this formulation explicitly retains the energy integral over the density of states, making it more suitable for systems with sharp peaks at the Fermi energy. These coupled equations are solved self-consistently to determine the electron mass renormalization $Z(i\omega_j)$, the energy shift $\chi(i\omega_j)$ and the order parameter $\phi(\varepsilon, i\omega_j) =   \phi^{\mathrm{ph}}(i\omega_j) +  \phi^{\mathrm{c}}(\varepsilon)$,

\begin{equation}
\begin{split}
    Z(i\omega_j) =   1+ &\frac{k_{\mathrm{B}} T}{N_{\rm{F}} \,\omega_j} \int d\varepsilon' \,N(\varepsilon') \\& \times\sum_{j'} \frac{\omega_{j'} Z(i\omega_{j'})}{\Theta(\varepsilon', i\omega_{j'})}\, \lambda(\omega_j - \omega_{j'})
\end{split}
    \label{eqn:Z}
\end{equation}

\begin{equation}
\begin{split}
\chi(i\omega_j) = &   -\frac{k_{\rm{B}}T}{N_{\rm{F}}} \int d\varepsilon'\,N(\varepsilon') \\& \times\sum_{j'} \frac{\varepsilon'-\mu_{\rm{F}} + \chi(i\omega_{j'})}{\Theta(\varepsilon', i\omega_{j'})} \lambda(\omega_j - \omega_{j'})
    \end{split}
    \label{eqn:chi}
\end{equation}


\begin{equation}
\begin{split}
    \phi^{\mathrm{ph}}(i\omega_j) = &\frac{k_{\rm{B}}T}{N_{\rm{F}}} \int d\varepsilon'\,N(\varepsilon') \\& \times\sum_{j'} \frac{\phi(\varepsilon', i\omega_{j'})}{\Theta(\varepsilon', i\omega_{j'})}\, \lambda(\omega_j - \omega_{j'})
    \end{split}
    \label{eqn:phi_ph}
\end{equation}

\begin{equation}
\begin{split}
    \phi^{\mathrm{c}}(\varepsilon) = &- k_{\rm{B}}T  \int d\varepsilon'\,N(\varepsilon') \,W(\varepsilon,\varepsilon') \\& \times \sum_{j'} \frac{\phi(\varepsilon', i\omega_{j'})}{\Theta(\varepsilon', i\omega_{j'})}
\end{split}
    \label{eqn:phi_c}
\end{equation}

where the denominator $\Theta(\varepsilon, i\omega_{j}) = (\omega_jZ(i\omega_j))^2 + (\varepsilon - \mu_F+\chi(i\omega_j))^2 + \phi(\varepsilon, i\omega_j)^2$. Here $\omega_j = (2j+1)\pi k_{\rm{B}}T$ are fermion Matsubara frequencies, $N_{\rm{F}}$ is the density of states at the Fermi energy and $N(\varepsilon')$ is the density of states at electronic energy $\varepsilon'$. The chemical potential $\mu_{\rm{F}}$ is determined self-consistently using an equation fixing the number of electrons
given in Ref.~\cite{lucreziFullbandwidthAnisotropicMigdalEliashberg2024a, koglerIsoMEStreamliningHighprecision2025a}. The term $\lambda(\omega_j - \omega_{j'})$ is the electron-phonon coupling kernel and is given by
\begin{equation}
    \lambda(\omega_j - \omega_{j'}) = \int_0^\infty d\omega \; \alpha^2F(\omega) \frac{2\omega}{(\omega_j - \omega_{j'})^2 + \omega^2}
\end{equation}
where $\alpha^2F(\omega)$ is the isotropic Eliashberg spectral function, given by
\begin{equation}
    \begin{split}
    \alpha^2F(\omega) = \frac{1}{N_{\mathrm{F}}}& \sum_{\mathbf{k}, \mathbf{k'},n,m} \sum_{\nu}  \;\left| g_{nm\nu}(\mathbf{k},\mathbf{k'})\right|^2  \\
    &\quad \times \,\delta(\varepsilon_{\mathbf{k},n}) \delta(\varepsilon_{\mathbf{k'},m}) \delta(\omega - \omega_{\mathbf{k} - \mathbf{k'}, \nu}).
    \end{split}
    \label{eqn:alpha2F-long}
\end{equation}
with $g_{nm\nu}(\mathbf{k},\mathbf{k'})$ being the electron-phonon matrix elements and $\omega_{\mathbf{k} - \mathbf{k'}, \nu}$ the phonon frequency for wavevector $\mathbf{q} = \mathbf{k-k'}$ and mode $\nu$. The $\delta(\varepsilon_{\mathbf{k}n})$ restricts the summation to electronic states at the Fermi level.

$W(\varepsilon,\varepsilon')$ is the constant-energy surface average of the screened Coulomb interaction between Bloch states $W_{\mathbf{k},n , \mathbf{k+q},m} $ given in Eq.~\eqref{eqn:w_nkmp}, and is defined as
\begin{equation}
\begin{split}
    W(\varepsilon,\varepsilon') = &\sum_{\mathbf{k},n , \mathbf{q},m} W_{\mathbf{k},n , \mathbf{k+q},m} \\& \quad\quad\quad \times\frac{\delta(\varepsilon - \varepsilon_{\mathbf{k},n})}{N(\varepsilon)} \frac{\delta(\varepsilon' - \varepsilon_{\mathbf{k+q},m})}{N(\varepsilon')}
    \end{split}
    \label{eq:W_ee}
\end{equation}

The dimensionless Coulomb parameter 
\begin{equation}
    \mu = N_{\rm{F}}\, W(\varepsilon_{\rm{F}}, \varepsilon_{\rm{F}}) 
\end{equation}
which is analogous to $\lambda$, characterizes the strength of the screened electron-electron interaction at the Fermi surface. For computational efficiency, it is common to introduce a Matsubara cutoff energy $\omega_{c}$, truncating the summation over Matsubara frequencies $\omega_j$. Above $\omega_c$, the phonons are not expected to contribute, allowing the introduction of approximations that mean the summations in \cref{eqn:Z,eqn:chi,eqn:phi_ph,eqn:phi_c} can be performed analytically. After this process, $\mu$ is replaced by $\mu^*$, the renormalized quantity, given by
\begin{equation}
    \mu^*_{\rm{ME}}  = \frac{\mu}{1 + \mu \ln(\frac{E_{\rm{el}}}{\omega_c})},
\end{equation}
with cutoff energy $\omega_c$ and electronic energy $E_{\rm{el}}$. However, as the value of $\mu^*$ depends on the choice of cutoff $\omega_c$, we have instead elected to report $\mu$ values as these are fully transferable between calculations. 

\section{Method and Computational Details}\label{sec:computation}

The \textsc{Mantel} code~\cite{mantel-v1} was developed to evaluate the static screened Coulomb matrix elements given in Eq.~\eqref{eqn:w_nkmp}, using the Bloch wavefunctions and static inverse dielectric from the \textsc{Quantum ESPRESSO} \cite{giannozziQUANTUMESPRESSOModular2009} and \textsc{Yambo} \cite{marini2009, sangalliManybodyPerturbationTheory2019} codes, respectively.

The starting point of the workflow is the DFT ground state calculated with \textsc{Quantum ESPRESSO}, using the PseudoDojo \textsc{oncvpsp} PBE pseudopotentials~\cite{hamannOptimizedNormconservingVanderbilt2013a, vansettenPseudoDojoTrainingGrading2018}. To determine the kinetic-energy cutoff, the total energy was converged to within 1~meV/atom, yielding cutoffs of around 100~Ry. The electronic $\mathbf{k}$-point spacing was set to $2\pi \times 0.03$~\AA$^{-1}$. These DFT wavefunctions were used as input to the \textsc{Yambo} code, which evaluates the static inverse dielectric using Eq.~\eqref{eqn:inv_dielec}. This involves evaluating the non-interacting irreducible polarizability, which, as shown in Eq.~\eqref{eqn:irred_polar}, requires a summation over unoccupied states.  Convergence tests suggested 100 total bands was suitable for most elemental and hydride systems, although for Mg$_2$IrH$_6$ a total of 120 bands were used. A cutoff of 15~Ry was used to determine the size of the $\epsilon^{-1}_{\mathbf{G},\mathbf{G'}}(\mathbf{q},0)$ matrix, and the $\mathbf{q}$-point grid was set to a spacing of $2\pi \times 0.06$~\AA$^{-1}$. Once the static inverse dielectric was calculated, it was given as input to the \textsc{Mantel} code. The \textsc{Quantum ESPRESSO} wavefunctions were truncated and padded to fit on a $17\times17\times17$ grid in reciprocal space, which was verified to be well converged for the systems tested. For the $\mathbf{q}\to0$ limit, the head of the screened interaction was replaced with the Thomas-Fermi value, and the wings were set to zero to handle the divergence of the bare Coulomb interaction at $\mathbf{q}=0$ and reflect the metallic screening.

The Eliashberg spectral function $\alpha^2F(\omega)$, which is required to calculate $T_{\mathrm{c}}$, was computed using density functional perturbation theory (DFPT) in \textsc{Quantum ESPRESSO}. The phonons were evaluated on a $\mathbf{q}$ grid set to a spacing of $2\pi\times0.06$~\AA$^{-1}$, and the coarse electronic $\mathbf{k}$ grid for the phonon calculation was set to $2\pi\times0.03$~\AA$^{-1}$. A fine electronic grid of twice this density was used to evaluate the electron-phonon matrix elements $g_{nm\nu}(\mathbf{k},\mathbf{k'})$. The calculation of the spectral function in Eq.~\eqref{eqn:alpha2F-long} involves delta functions $\delta(\varepsilon_{\mathbf{k},n})$ and $\delta(\varepsilon_{\mathbf{k'},m})$, which in practice are replaced by Gaussian functions of smearing width $\sigma$. To mitigate the sensitivity to the smearing width, we used the density of states rescaling technique~\cite{bozierHighthroughputSuperconducting$T_c$2026}, using a tetrahedron density of states evaluated on an electronic $\mathbf{k}$ grid with a spacing of $2\pi \times 0.02$~\AA$^{-1}$.

To calculate $T_{\mathrm{c}}$ using the variable-DOS Eliashberg equations shown in \cref{eqn:Z,eqn:chi,eqn:phi_ph,eqn:phi_c}, we utilized the \textsc{IsoME} code~\cite{koglerIsoMEStreamliningHighprecision2025a}. The constant-energy surface average in Eq.~\eqref{eq:W_ee} was evaluated by replacing the delta functions with Gaussian functions of width 0.3~eV. A Matsubara cutoff frequency of 7000~meV was used for H$_3$S, while Mg$_2$IrH$_6$  used a cutoff of 2000~meV. To calculate $\mu = N_{\mathrm{F}}\, W(\varepsilon_{\mathrm{F}}, \varepsilon_{\mathrm{F}})$, which enables direct comparison of the interaction strength between systems, we used the tetrahedron density of states at the Fermi energy $\widetilde{N}_{\mathrm{F}}$ as this converged more rapidly than a Gaussian-smeared DOS. When solving the standard $\mu^*$ form of the Eliashberg equations, rather than the fully energy-dependent $W(\varepsilon, \varepsilon')$ form, the typical electronic energy scale entering the renormalization from $\mu$ to $\mu^*$ was estimated from the bandwidth of bands crossing the Fermi level~\cite{heil2019}.

Convergence tests demonstrating how the calculated $\mu$ value varied with the wavefunction cutoff, the dielectric cutoff energy, the number of empty bands, the $\mathbf{q}$ grid spacing, the $\mathbf{k}$ grid spacing, and the \textsc{Mantel} wavefunction grid size are shown in the Supplemental Material.

\section{Results}\label{sec:results}

\subsection{Benchmark of elements}\label{sec:benchmark_elements}

\begin{figure}[ht!]
    \centering
    \includegraphics[width=0.95\linewidth]{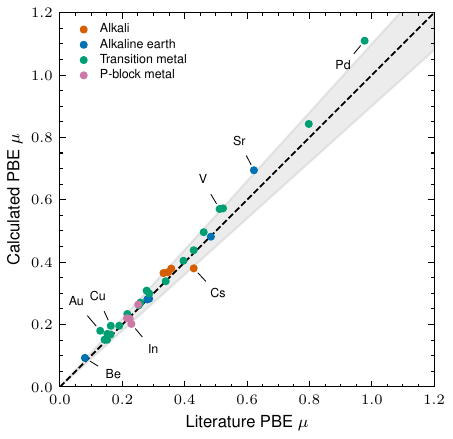}
    \caption{Parity plot for the ambient pressure metallic elements, comparing the $\mu$ value calculated with the PBE functional in \textsc{Mantel} to those obtained in Ref.~\cite{kawamuraBenchmarkDensityFunctional2020}. The grey region indicates a fractional difference within $\pm10$\%.}
    \label{fig:parity}
\end{figure}

To validate the \textsc{Mantel} predictions, we calculated the value of $\mu$ for the non-magnetic metallic elements. The obtained values were then compared to a benchmark study by Kawamura et al.~\cite{kawamuraBenchmarkDensityFunctional2020}, which investigated the screened Coulomb interaction strength relevant for superconducting density functional theory (SCDFT). Although our results are used instead within the Eliashberg equations, the static Fermi-surface averaged quantity evaluated in both cases ($\mu$) is the same. Both sets of values are calculated using the PBE functional. A parity plot of the results is shown in Fig.~\ref{fig:parity}, with the shaded region indicating a fractional difference within $\pm10$\%. Good agreement is achieved overall, with the majority of elements sitting within the shaded region. Interestingly, both the alkali and \textit{p}-block metals display minimal spread, compared to the alkaline earth metals where the value of $\mu$ for Sr (0.695) is several times larger than that of Be (0.093). The transition metals also display large spread, but this is expected given that different fillings of the \textit{d}-shell lead to different screening behaviour, density of states, and orbital character at the Fermi level. For the majority of materials, there appears to be a slight tendency to predict a larger $\mu$ than in Ref.~\cite{kawamuraBenchmarkDensityFunctional2020}, which may arise from differences in the Brillouin-zone integration scheme.

\begin{figure}[h!]
    \centering
    \includegraphics[width=\linewidth]{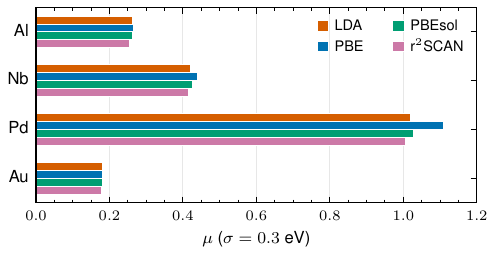}
    \caption{Comparison of impact of exchange-correlation potential on the $\mu$ value. The values are largely insensitive to the choice of functional, with discrepancies driven by changes in the electronic density of states at the Fermi level.}
    \label{fig:xc_compare}
\end{figure}

We next investigated the sensitivity of $\mu$ to the exchange-correlation functional for Al, Au, Nb, and Pd, as shown in Fig. \ref{fig:xc_compare}. We evaluated $\mu$ using the LDA, PBE, and PBEsol functionals. We also included the r$^2$SCAN functional. However, because the dielectric could not be computed directly within this framework, the screening was approximated using the PBE dielectric function, while r$^2$SCAN was used to determine the equilibrium volume and wavefunction overlaps. In principle, the choice of functional can impact $\mu$ in several ways, through differences in the electronic density of states, the shape and overlap of wavefunctions, and the starting point of the polarizability calculation. The latter depends on the non-interacting Green's functions constructed from the electronic Kohn-Sham eigenvalues. In practice, however, the results shown in Fig.~\ref{fig:xc_compare} demonstrate that in most cases the choice of functional has little effect on $\mu$. The minor discrepancies between the functionals arise from changes in the electronic density of states, rather than the strength of the underlying screened Coulomb interaction.

\begin{figure*}[ht]
    \centering
    \includegraphics[width=\linewidth]{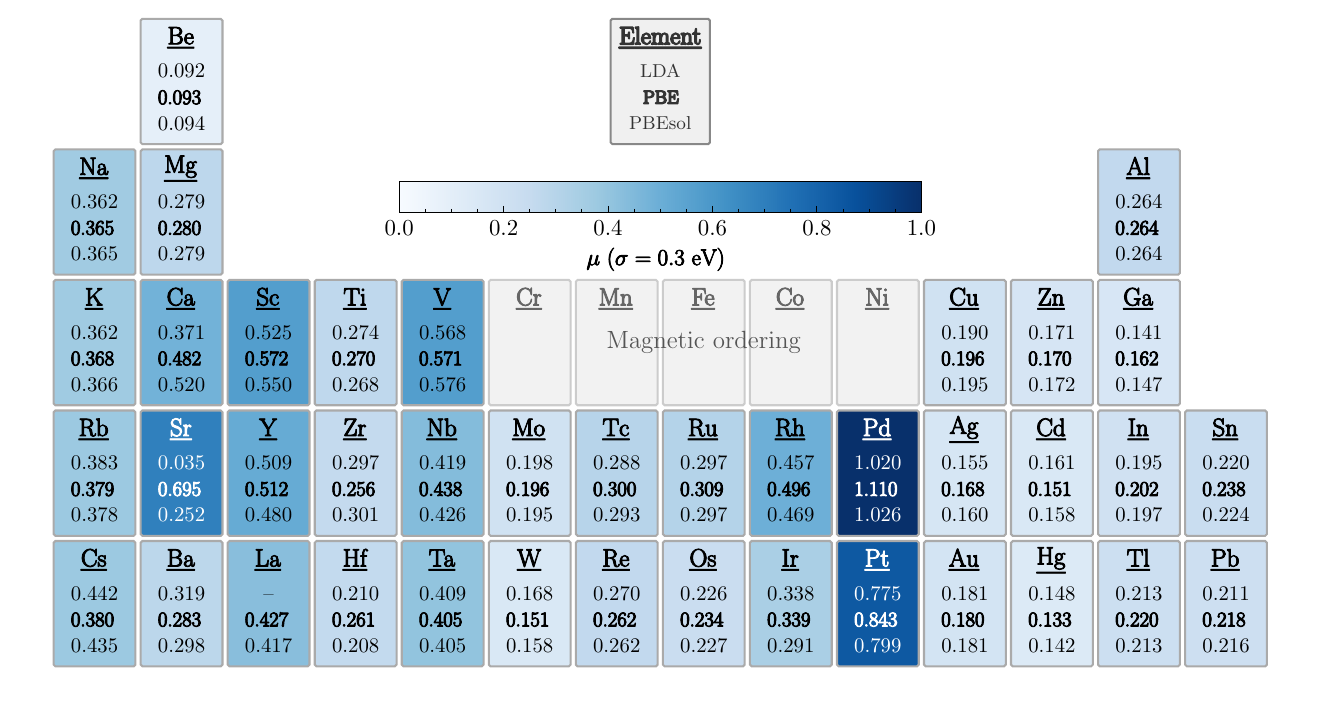}
    \caption{Periodic table of the ambient pressure metallic elements, with the value of Coulomb parameter $\mu$ given with the LDA, PBE and PBEsol functionals. The colour is set by the PBE value, shown in bold. The elements Cr to Ni are excluded due to their formation of magnetic ordering, which typically precludes conventional superconductivity. The LDA for La is omitted due to the lack of a \textsc{pseudo-dojo} pseudopotential.}
    \label{fig:periodic_table}
\end{figure*}

The LDA, PBE, and PBEsol numerical results are presented in the periodic table in Fig.~\ref{fig:periodic_table}. The elements are coloured by the PBE value. Several trends emerge. First, Group 1 metals all display closely clustered values centred around 0.38, which we attribute to the very similar $N_{\mathrm{F}}$ values and the same bcc crystal structure down the group. In contrast, the Group 2 metals adopt a range of structures: hcp for Be and Mg, fcc for Ca and Sr, and bcc for Ba. This produces a wide range of $N_{\mathrm{F}}$ values, resulting in both the smallest (Be) and third-largest (Sr) Coulomb parameter values. The large spread in Sr values between functionals is directly driven by its rapidly-varying density of states, which results in very different $N_{\mathrm{F}}$ values for small shifts in the Fermi level. Moving to the transition metals, the Pd and Pt systems stand out for their very large $\mu$ values, which we attribute to the large $d$-electron peak near the Fermi energy. This peak leads to a large $N_{\mathrm{F}}$, and since the $d$-electrons tend to screen less effectively than the $s$- or $p$-electrons, we also expect a stronger Coulomb interaction. Lastly, the late-transition-metal and $p$-block elements have relatively uniform $\mu$ values, typically lower than those of the $s$-block elements. We note that the density of states alone is not sufficient to fully predict trends in $\mu$ values. For example, Al and Re possess similar $\mu$ values despite the latter having a significantly larger $N_{\mathrm{F}}$ (0.78 states/eV/atom) than the former (0.41 states/eV/atom).

\subsection{Common Superconductors}\label{sec:common}
\begin{figure}
    \centering
    \includegraphics[width=\linewidth]{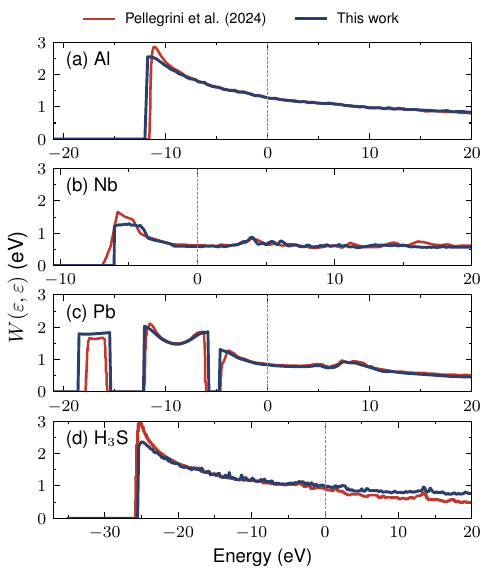}
    \caption{Plots of the diagonal of the isoenergy averaged screened Coulomb interaction $W(\varepsilon, \varepsilon)$. Data from Ref.~\cite{pellegriniInitioMethodsSuperconductivity2024} is shown in red and from this work in blue. The Fermi level is denoted by the dashed line. Note the different energy ranges for (b) and (d).}
    \label{fig:sanna_compare}
\end{figure}

We calculated the screened Coulomb interaction $W$ for several superconducting systems and compared it to the work of Pellegrini and Sanna in Ref.~\cite{pellegriniInitioMethodsSuperconductivity2024}. Plots of the diagonal of the isoenergy average, $W(\varepsilon, \varepsilon)$, are shown in Fig.~\ref{fig:sanna_compare}. The red and blue curves denote the Ref.~\cite{pellegriniInitioMethodsSuperconductivity2024} results and the \textsc{Mantel} results, respectively. In regions of energy with no electronic states, $W(\varepsilon, \varepsilon)$ is set to zero. In contrast to the dimensionless Coulomb parameter $\mu$, care must be taken when comparing $W$ values, as the value depends on the choice of simulation cell. Throughout this paper, we use the primitive unit cell containing one formula unit. 

The agreement is generally very good, particularly in regions where the interaction is smooth and varies slowly with energy. Discrepancies are seen at the sharp peaks. We attribute this to sensitivities in the density of states. As the isoenergy average includes factors of $\frac{1}{N(\varepsilon)}$, when the density of states becomes small this can lead to large differences. Perhaps surprisingly, niobium has a large value of $\mu$, yet the screened Coulomb interaction at the Fermi energy is noticeably weaker than in Al. This is due to its high density of states at the Fermi level. Although a large $N_{\mathrm{F}}$ is beneficial for a sizeable Coulomb parameter $\mu$ due to the linear $N_{\mathrm{F}}$ prefactor, it also enhances the strength of the polarizability $\chi^0$ and hence improves screening, leading to a weaker Coulomb interaction. Put simply, a higher density of states at the Fermi energy means the system is more metallic, so charges are screened more effectively. On balance however, this effect is not enough to cancel out the linear prefactor, and so a system with a large density of states will still typically have a large $\mu$ value. This $N_{\mathrm{F}}$ prefactor accounts for the fact that if more states are available to scatter into and participate in the interaction, then the overall interaction will be stronger. The same reasoning is used to explain why a large $N_{\mathrm{F}}$ is beneficial for high electron-phonon coupling strength $\lambda$.

Both the Al and H$_3$S systems display smooth decays as a function of energy. This behaviour is characteristic of a uniform electron gas and is found in systems with relatively homogeneous charge densities. In the homogeneous electron gas, this decay arises because at higher energies, larger $\mathbf{q}$ scatterings become possible. These large $\mathbf{q}$ scatterings both dominate the phase space and have smaller $ W_{\mathbf{k},n , \mathbf{k+q},m} $ matrix elements. In comparison, systems with a more inhomogeneous electron density display more variation as a function of energy, such as in Nb where there is a sharp drop from the peak centred around $-5$ eV, and a subsequent minor peak at 4 eV. Likewise, Pb shows interesting behaviour, with gaps in the density of states leading to regions where $W$ drops to zero. In general, the more homogeneous the electron density, the smoother the observed $W(\varepsilon, \varepsilon)$ plot.

\subsection{Hydride Superconductors}\label{sec:hydrides}
\begin{figure*}
    \centering
    \includegraphics[width=\linewidth]{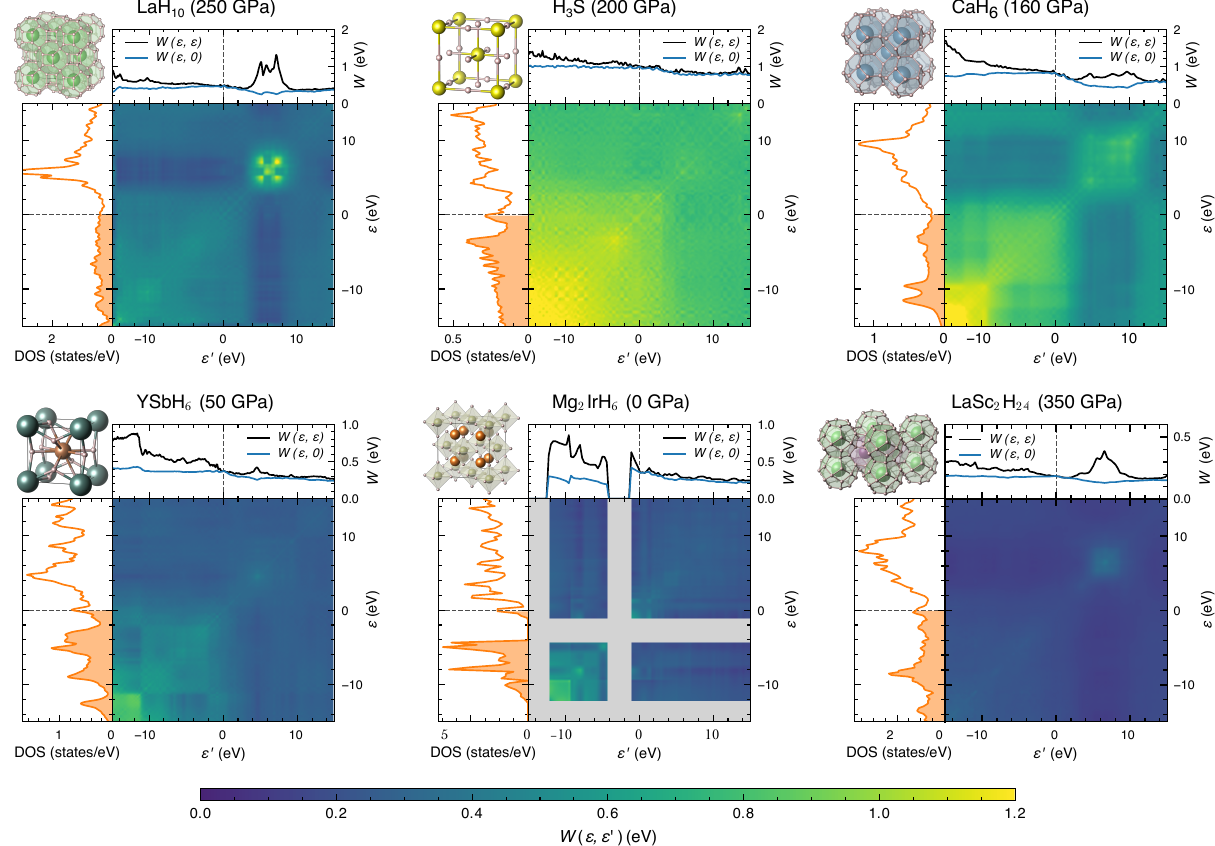}
    \caption{Comparison of a range of superconducting hydride systems. Each subfigure shows the density of states, the diagonal $W(\varepsilon, \varepsilon)$ and horizontal $W(\varepsilon,0)$ cross sections, and a colourplot of the full matrix. The same colour range is used across all systems, illustrating the screened Coulomb matrix elements are strong for H$_3$S, and weak for LaSc$_2$H$_{24}$.}
    \label{fig:hydrides}
\end{figure*}

The hydrides are a promising class of conventional superconductors, with the light mass of hydrogen leading to high average phonon frequencies and high $T_{\mathrm{c}}$. However, many of these systems only form at high pressures, and it is unclear whether the standard values of $\mu^*$ are suitable. As such, we investigated several hydrides spanning a wide pressure range, including experimentally verified LaH$_{10}$~\cite{drozdovSuperconductivity250Lanthanum2019}, H$_3$S~\cite{drozdovConventionalSuperconductivity2032015b}, CaH$_6$~\cite{liSuperconductivity200Discovered2022}, and predicted systems YSbH$_6$~\cite{causseMetastabilityHigh$T_c$Superconductivity2026}, Mg$_2$IrH$_6$~\cite{doluiFeasibleRouteHighTemperature2024, sannaPredictionAmbientPressure2024} and LaSc$_2$H$_{24}$~\cite{he2024, song2025, wang2026, semenok2026, semenok2026a}. Figure~\ref{fig:hydrides} shows the density of states (orange), a colourmap of the screened Coulomb isoenergy average $W(\varepsilon, \varepsilon')$, and the diagonal ($\varepsilon=\varepsilon'$) and Fermi energy ($\varepsilon'=0$) cross sections of this matrix. 

We first consider the magnitude of the screened Coulomb interaction, with the values here given per formula unit. Figure \ref{fig:hydrides}  shows a spread of maximum magnitudes, ranging from above 1 eV down to only 0.35 eV for LaSc$_2$H$_{24}$. Again, the key variable behind this is the density of states at the Fermi energy $N_{\mathrm{F}}$. A larger $N_{\mathrm{F}}$ directly leads to a larger polarizability $\chi^0$ and better screening. As seen in Eq.~\eqref{eqn:irred_polar}, the states at the Fermi level contribute the most because the polarizability requires a transition from a filled to an empty state, weighted by the inverse of the energy difference between the states. Comparing values of $N_{\mathrm{F}}$ across systems, we see a clear trend with the more metallic systems exhibiting smaller $W$ magnitudes. However, while large $N_{\mathrm{F}}$ does suppress $W$, the product $\mu = N_{\mathrm{F}}W(\varepsilon_{\mathrm{F}}, \varepsilon_{\mathrm{F}})$ generally still increases.

Several of the systems display a peak in the diagonal of $W$ at energies corresponding to a peak in the density of states. These include LaH$_{10}$ with its sharp peak at 6 eV, and CaH$_{6}$ with a broader peak between 4 eV and 12 eV. We attribute these features to strongly localized, degenerate $f$-like and $d$-like states which share the same centre. Such states exhibit strong spatial overlap, leading to large $\rho_{n\mathbf{k}, m\mathbf{k+q}}(\mathbf{G})$ terms and consequently large $W_{n\mathbf{k}, m\mathbf{k+q}}$ matrix elements. However, being several eV above the Fermi level, these orbitals do not significantly contribute to the polarizability, and thus do not greatly increase the screening for LaH$_{10}$. The calculated values of $\mu=0.18$ for LaH$_{10}$ and $\mu=0.14$ for CaH$_{6}$ are low, and thus the standard $\mu^*=0.1$--$0.15$ is appropriate for these systems once the renormalization is applied.

The H$_3$S system, as discussed previously, displays a smooth decay in $W$. The $\varepsilon$ and $\varepsilon'$ corresponding to low-energy states, with the lower density of states, have a larger value of $W(\varepsilon,\varepsilon')$. This is because they typically involve smaller $\mathbf{q}$ scatterings, for which the bare and screened Coulomb interactions are stronger. While well known for its van Hove singularity near the Fermi energy, the actual $N_{\mathrm{F}}$ value, expressed in states/eV/f.u., is smaller than in most other hydrides, and thus leads to large $W$. Nonetheless, the product $\mu = N_{\mathrm{F}} W(\varepsilon_{\mathrm{F}}, \varepsilon_{\mathrm{F}})$ remains moderately large at $\mu=0.30$, and consequently $T_{\mathrm{c}}$ is somewhat suppressed compared to the standard $\mu^*$ values.

For the remaining systems with large $N_{\mathrm{F}}$ values, we expect large $\mu$ values. The YSbH$_6$ system follows this trend, yielding a large $\mu=0.33$. The strong Coulomb interactions consequently suppress $T_{\mathrm{c}}$ from the 118~K reported in Ref.~\cite{causseMetastabilityHigh$T_c$Superconductivity2026} down to around $100$~K. In contrast, LaSc$_2$H$_{24}$ is something of an outlier. Despite having the second largest $N_{\mathrm{F}}$ of this set, behind only Mg$_2$IrH$_6$, the obtained $\mu=0.24$ is lower than might be expected and lies between LaH$_{10}$ and H$_3$S. One possible explanation is that the broad peak in the DOS just above the Fermi energy results in improved screening and a smaller inverse dielectric.

The Mg$_2$IrH$_6$ system features energy regions with no electronic states, which consequently have $W=0$.  It has been argued that such features, which narrow the bandwidth, impact the renormalization procedure of going from $\mu$ to $\mu^*$ by reducing the effective electronic energy scale~\cite{koch1999, albertoummarino2026}. These features also tend to produce peaks in the diagonal $W$ immediately adjacent to them. This is likely because only a few $\mathbf{k}$ states are available to couple, and those that do are connected by small $\mathbf{q}$ scatterings. While the magnitude of $W$ is suppressed by the large $N_{\mathrm{F}}$, the obtained $\mu=0.62$ is exceptionally large, which we attribute to both the large $N_{\mathrm{F}}$ and the energy gaps leading to these peaks in $W$. This supports the hypothesis presented in Ref.~\cite{sannaPredictionAmbientPressure2024}, that the unusual electronic structure of Mg$_2$IrH$_6$ means the standard $\mu^*_{\text{ME}}=0.16$ commonly used for this system is not appropriate. For example, using a Matsubara cutoff frequency of 2000~meV and a bandwidth of 5~eV, we obtain $\mu^*_{\text{ME}}\approx0.39$, substantially larger than the $\mu^*_{\text{ME}}=0.16$ previously used.

Overall, the behaviour of $W$ is complex, and closely depends on features of the underlying electronic structure. This underscores the necessity of explicit first-principles calculations. As a general trend, a large $N_{\mathrm{F}}$ suppresses the magnitude of $W$, but usually yields a net increase in $\mu$. Localised electronic orbitals, which typically cause sharp peaks in the DOS, can produce peaks in the $W$ values, due to strong spatial and reciprocal space overlaps. Conversely though, low DOS regions can also result in large $W$ if the allowed states are connected by small scattering vectors $\mathbf{q}$. This is because scattering with small $\mathbf{q}$ is where both the bare and screened Coulomb interactions are strongest. The broad spread of obtained $\mu$ values, ranging from 0.14 to 0.62, illustrates the importance of being able to calculate this parameter from first-principles.

\subsection{Pressure Dependence}\label{sec:pressure}
\begin{figure}[h!]
    \centering
    \includegraphics[width=\linewidth]{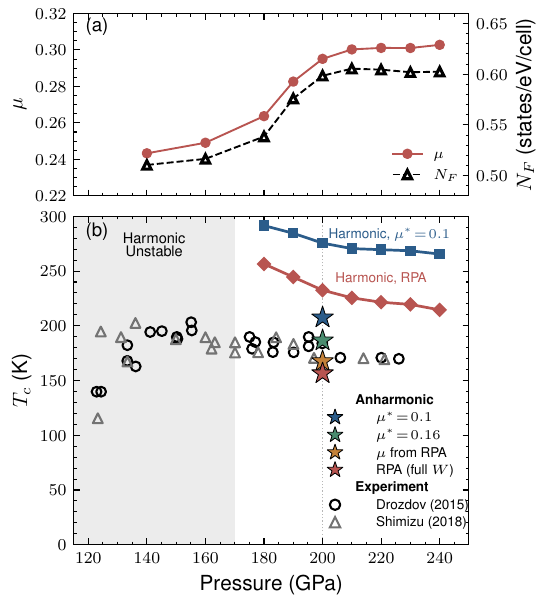}
    \caption{(a) Pressure dependence of $\mu$ and $N_{\mathrm{F}}$ for H$_3$S. (b) Resulting pressure evolution of $T_{\mathrm{c}}$ for H$_3$S. The anharmonic $\alpha^2F(\omega)$ at 200~GPa was taken from Ref.~\cite{errea2015}, and used to solve the variable-DOS Eliashberg equations with different strength Coulomb interactions. The dotted line at 200~GPa is a guide to the eye.}
    \label{fig:pressure_H3S}
\end{figure}

Predictions of $T_{\mathrm{c}}$ under pressure typically assume a fixed $\mu^*$, neglecting any pressure dependence of the Coulomb interaction. We therefore examined whether this is justified. Figure~\ref{fig:pressure_H3S} shows the results for H$_3$S, with panel (a) illustrating how both $\mu$ and the density of states at the Fermi energy $N_{\mathrm{F}}$ evolve in the pressure range 140~GPa to 240~GPa. The $\mu$ parameter follows the trends in $N_{\mathrm{F}}$, suggesting that the isoenergy-averaged screened Coulomb interaction $W$ is largely insensitive to pressure within this system. The changes in $\mu$ are comparatively small, with a moderate increase from 0.24 to 0.3, and are typically smaller than those seen in the electron-phonon coupling strength $\lambda$, which drops from 2.7 at 200~GPa to 2.1 at 240~GPa. We attribute this much larger change to the fact that $\lambda$ also depends on phonon frequencies, which can vary strongly with lattice parameter. Consequently, most variation in $T_{\mathrm{c}}$ as a function of pressure comes from the electron-phonon interaction. 

Nonetheless, the use of the first-principles Coulomb interaction has a notable impact on the magnitude, and a minor impact on the trends for $T_{\mathrm{c}}$, as shown in Fig.~\ref{fig:pressure_H3S} (b). All calculated $T_{\mathrm{c}}$ values here employ the variable DOS Eliashberg equations. The red curve, which denotes the $T_{\mathrm{c}}$ obtained including Coulomb effects from \textsc{Mantel}, sits 35 to 50~K below the fixed $\mu^*=0.1$ blue curve, and brings the $T_{\mathrm{c}}$ estimate closer to the experimental values. Additionally, a minor difference in trends is observed, as the increase in $\mu$ with pressure leads to a stronger suppression of $T_{\mathrm{c}}$ in the red curve at higher pressures.

As anharmonicity has been found to be important in the high-pressure hydrides~\cite{errea2013,errea2015, hou2021, belli2026}, we also consider the $T_{\mathrm{c}}$ obtained using anharmonic phonon frequencies. Anharmonicity modifies the phonon frequencies and hence electron-phonon coupling strength, and alters the pressure range over which the structure remains stable. For H$_3$S, the harmonic approximation leads to unstable phonons below around 170~GPa, shown by the grey shaded region, despite the superconducting $Im\overline{3}m$ structure being experimentally observed well below this pressure. In contrast, inclusion of anharmonicity has been shown to stabilise the structure down to around 100~GPa~\cite{belli2026}.

The anharmonic Eliashberg spectral function was taken from Ref.~\cite{errea2015} and used to obtain the anharmonic $T_{\mathrm{c}}$ for different Coulomb interactions. Data from experimental measurements are shown with open circles and triangles~\cite{drozdovConventionalSuperconductivity2032015b,shimizu2018}. The use of the anharmonic phonons substantially lowers $T_{\mathrm{c}}$ by around 70~K, driven by the drop in $\lambda$ from 2.7 for the harmonic to 1.8 for the anharmonic. The blue and green points show the results for fixed $\mu^*=0.1$ and $\mu^*=0.16$ respectively, with the latter being widely used in the literature as it leads to the best agreement with the experimental data from Drozdov et al.~\cite{drozdovConventionalSuperconductivity2032015b}. In comparison, the first-principles calculations in orange and red sit slightly lower, although still in quite good agreement, especially with the data from Shimizu et al.~\cite{shimizu2018}. Interestingly, the nominally highest level of theory shown in red now underestimates the $T_{\mathrm{c}}$ by 10 to 20~K, with the Fermi-surface averaged $\mu$ approach (orange) giving better agreement than the energy-dependent $W$ treatment (red). Overall, both the \textit{ab initio} Coulomb interaction and inclusion of anharmonicity have a sizeable impact on the predicted $T_{\mathrm{c}}$, with the first-principles treatments bringing the estimate into good agreement with experiment.

\begin{figure}[h]
    \centering
    \includegraphics[width=\linewidth]{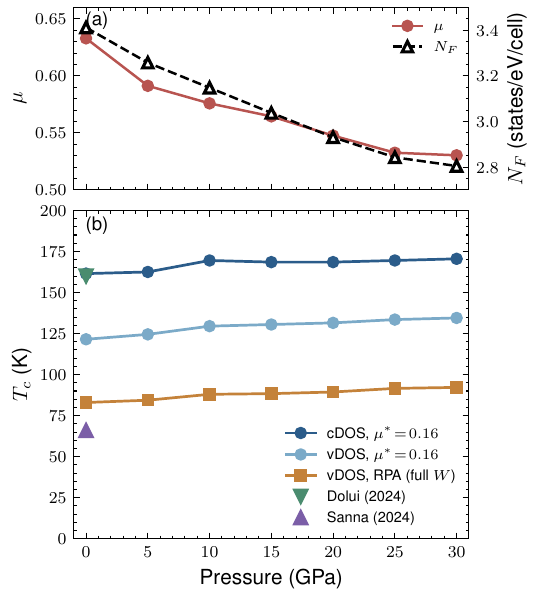}
    \caption{(a) Pressure dependence of $\mu$ and $N_{\mathrm{F}}$ for Mg$_2$IrH$_6$. (b) Resulting pressure evolutions of $T_{\mathrm{c}}$ for Mg$_2$IrH$_6$. Compared to the constant DOS (cDOS) and $\mu^*$ approach, using the variable DOS (vDOS) and first principles Coulomb interactions (full $W$) substantially suppresses the $T_{\mathrm{c}}$. }
    \label{fig:press_Mg2IrH6}
\end{figure}

Figure~\ref{fig:press_Mg2IrH6} shows a similar plot for Mg$_2$IrH$_6$. Again, in panel~(a) $\mu$ tracks $N_{\mathrm{F}}$, and this time decreases with pressure. The resulting $T_{\mathrm{c}}$ in (b) shows only a very minor increase with pressure. The two theoretical predictions \cite{doluiFeasibleRouteHighTemperature2024, sannaPredictionAmbientPressure2024} at ambient pressure are shown with triangles; this system has not been synthesized and so no experimental data exist. The dark blue curve, representing the constant DOS and $\mu^*$ approach, is the standard approach used in the literature, and our results agree perfectly with Dolui et al.~\cite{doluiFeasibleRouteHighTemperature2024}. When the variable DOS approach is used, the $T_{\mathrm{c}}$ drops by around 40~K as seen in the light blue. This is due to the sharp peak in the density of states at $\varepsilon_{\mathrm{F}}$, which makes $N_{\mathrm{F}}$ unrepresentatively large compared to the DOS away from the Fermi level, and causes the constant DOS approach to overestimate. A further consequence of this large $N_{\mathrm{F}}$ is the large $\mu$ value, which leads to the red RPA curve being a further 40~K lower, in close agreement with the value from Sanna et al.~\cite{sannaPredictionAmbientPressure2024}. Consequently, the discrepancy between these two literature values can largely be explained by their respective choice of approximations: Ref.~\cite{doluiFeasibleRouteHighTemperature2024} employs the constant DOS approach with fixed $\mu^*$ while Ref.~\cite{sannaPredictionAmbientPressure2024} incorporates \textit{ab initio} Coulomb interactions and uses the variable DOS equations. Notably, Ref.~\cite{doluiFeasibleRouteHighTemperature2024} employs the approach most widely used across superconductivity predictions and is excellently converged within these approximations. However, our calculations suggest that for this particular system, the unusual density of states leads to very large Coulomb interactions and a suppression of $T_{\mathrm{c}}$, in agreement with the arguments of Ref.~\cite{sannaPredictionAmbientPressure2024}.  

\section{Discussion}\label{sec:discussion}

The full calculation of $\mu$ involves several convergence parameters, including the $\mathbf{q}$-point grid, the  $\mathbf{k}$-point grid, and the number of empty (or total) bands in the dielectric. Convergence plots for these quantities are given in the Supplemental Material. Table~\ref{tab:timings} shows the wall-time for several different calculations of bcc Nb, demonstrating that the number of $\mathbf{q}$- and $\mathbf{k}$-points is the dominant factor in the computational cost. For this simple system, however, the $\mu$ value converges rapidly and an accurate result can be obtained at relatively low cost. Indeed, this cost is lower than that of a typical phonon calculation, indicating that 
future high-throughput screening with the method is feasible. However, we note that the cost of a \textsc{Mantel} calculation scales steeply with system size, and for large unit cells, the computational expense may be significant.

\begin{table}
\caption{Computational cost breakdown and convergence of $\mu$ for 
Nb. Timings are reported in wall-time hours on an Intel(R) Xeon(R) 
Gold 6142 CPU @ 2.60\,GHz with 32 cores.}
\begin{ruledtabular}
\begin{tabular}{lcccc}
 & (i) & (ii) & (iii) & (iv) \\
\colrule
\multicolumn{5}{l}{\textit{Input parameters}} \\
\quad $\mathbf{q}$-points    & $6^3$  & $12^3$ & $12^3$ & $12^3$ \\
\quad $\mathbf{k}$-points    & $12^3$ & $12^3$ & $12^3$ & $24^3$ \\
\quad \textsc{Yambo} bands   & 100    & 100    & 250    & 100    \\
\colrule
\multicolumn{5}{l}{\textit{Result}} \\
\quad $\mu$                  & 0.451  & 0.438  & 0.428  & 0.430  \\
\colrule
\multicolumn{5}{l}{\textit{Timings}} \\
\quad Inv.\ dielectric       & 0.03   & 0.38   & 0.74  & 0.35   \\
\quad Wavefunctions          & 0.18   & 0.18   & 0.18  & 0.90   \\
\quad \textsc{Mantel}        & 0.82   & 1.68   & 1.68 & 28.40  \\
\colrule
\quad Total                  & 1.03   & 2.24   & 2.60 & 29.65  \\
\end{tabular}
\end{ruledtabular}
\label{tab:timings}
\end{table}

The impact of first-principles Coulomb effects on the predicted $T_{\mathrm{c}}$ varies greatly between systems. For many elements, such as aluminium and lead, $T_{\mathrm{c}}$ is already well-predicted using standard $\mu^*$ values. Among the hydrides, LaH$_{10}$ and CaH$_{6}$ have very small $\mu$ values and a smoothly varying density of states around the Fermi level. For such systems, the standard $\mu^*$ approach already works well, and first-principles calculations confirm this. However, our results show several other systems have stronger Coulomb interactions than $\mu^*=0.1$--$0.15$ would suggest. Both YSbH$_6$ and H$_3$S have moderate $\mu \approx 0.3$, which notably suppresses $T_{\mathrm{c}}$ by around 20~K and 40~K respectively, while the $T_{\mathrm{c}}$ of Mg$_2$IrH$_6$ is suppressed further by its exceptionally large $\mu=0.62$. Given that reasonable estimates of $\mu$ can be obtained at low cost, we argue that first-principles $\mu$ calculations ought to become standard for future $T_{\mathrm{c}}$ estimates.

The behaviour of the screened Coulomb interaction and of $\mu$ is complex, and necessitates first-principles calculations. Nonetheless, our results show that certain features of the electronic structure can lead to stronger Coulomb interactions. First is the density of states at the Fermi energy. Large $N_{\mathrm{F}}$ values lead to a large $\mu$, despite also enhancing screening. Importantly, such systems are often among the most promising candidates for high $T_{\mathrm{c}}$ superconductivity, as the high $N_{\mathrm{F}}$ also promotes a large electron-phonon coupling $\lambda$. Thus, the systems where $\mu^*$ is least reliable are those of most interest for maximising $T_{\mathrm{c}}$. Systems where the Fermi level is near a bandgap are also susceptible to larger $\mu$, as the restrictions on allowed scatterings seem to favour large $W_{n\mathbf{k},m\mathbf{k+q}}$ matrix elements. The orbital character of states near the Fermi level also contributes to larger $\mu$, as localised $d$ and $f$ states possess strong intraband overlap while also screening the Coulomb interaction less efficiently. In the case of Mg$_2$IrH$_6$, all of these factors appear: the very large $N_{\mathrm{F}}$, the proximity of the large DOS peak to a bandgap, and a strong contribution from the Ir $d$ electrons, which together produce the large $\mu=0.62$.

Turning to the evolution with pressure, the screened Coulomb interaction $W$ for the systems investigated is relatively insensitive to pressure; nearly all the variation in $\mu$ is driven by changes in $N_{\mathrm{F}}$. For this reason, once a calculation of $\mu$ has been performed at one pressure, a rough estimate of $\mu$ at other pressures can be obtained by following the changes in $N_{\mathrm{F}}$. There does not appear to be a clear distinction between the magnitude of $\mu$ values obtained at high pressures and those at ambient conditions. Rather, what matters is whether the resulting electronic structure has the features identified above.

 One known limitation of the method as implemented is the use of the RPA dielectric, which corresponds to neglecting exchange-correlation effects in the polarizability. Pellegrini et al. \cite{pellegriniInitioCalculationsSuperconducting2023} investigated the impact of incorporating effective vertex corrections using the Kukkonen-Overhauser interaction, finding that their inclusion increased the calculated value of $\mu$. The discrepancy was smaller in the high electronic density limit ($r_s<1$), and grew progressively as the electron density decreased. These results are further corroborated by Ref.~\cite{cai2026}, which employed diagrammatic Monte Carlo to incorporate higher-order Feynman diagrams into the calculation of $\mu^*$ for the uniform electron gas. Although this analysis is limited to homogeneous densities, they likewise observed that as $r_s$ increased, corresponding to increasingly strong interactions and low density systems, the calculated $\mu$ was observed to be significantly greater than the RPA result. The impact was particularly stark for $r_s\geq2$, where the obtained $\mu$ exceeded twice the RPA value. Consequently, we expect the RPA approach to be most accurate for systems with high electron density; for low density systems, the current RPA estimation will underestimate $\mu$, and consequently overestimate the superconducting $T_{\mathrm{c}}$~\cite{rietschel1983}.

A further approximation is the use of the static dielectric function  $\epsilon(\mathbf{q}, 0)$ in place of the full dynamic function $\epsilon(\mathbf{q}, \omega)$. This corresponds to neglecting any frequency dependence of the screened Coulomb interaction $W$. In bulk metals, the static approximation is expected to perform well, as the plasmon energy scale ($\approx10$ eV) is significantly larger than the phonon energy scale~\cite{spataru2026}. However, a previous study by Davydov et al.~\cite{davydovInitioTheoryPlasmonic2020} quantified the impact of this frequency dependence on $T_{\mathrm{c}}$, finding a significant overestimation of predicted values. They observed that although static screening reliably reproduced experimental values to within 30\% on average, directly replacing the static dielectric function with the dynamic one led to average errors of 320\%. In their analysis, they highlighted the importance of the Coulomb contribution to the mass renormalization factor $Z^c$, the inclusion of which brought predictions closer to both static and experimental values. A later work by Akashi~\cite{akashi2022} showed that dynamic Coulomb interactions modify both the mass renormalization $Z^{c}$ and the energy shift $\chi^{c}$. These modifications cooperatively suppress $T_{c}$, further correcting the overestimates. Nevertheless, given the challenges and computational cost of including dynamic Coulomb interactions, we have elected to use the static approximation here.

\section{Conclusion}\label{sec:conclusion}

In summary, we have presented a first-principles study of the Coulomb parameter $\mu$ and Coulomb pseudopotential $\mu^*$ in superconducting hydrides across a range of pressures. Using the \textsc{Mantel} code, validated against elemental and common superconducting systems, we find a wide range in the strength of $\mu$, from 0.14 for CaH$_6$ to 0.62 for Mg$_2$IrH$_6$. In the latter case, the large Coulomb interactions suppress $T_{\mathrm{c}}$ by around 40~K, and resolve the discrepancy between previous theoretical estimates of $T_{\mathrm{c}}$. More broadly, the standard $\mu^*$ values adopted in the literature are often insufficient, failing to capture the strong materials dependence of the Coulomb interaction. We further show the pressure dependence of $\mu$ is largely governed by changes in the density of states. Given their significant impact on $T_{\mathrm{c}}$ demonstrated here, we suggest first-principles Coulomb interactions be incorporated as standard practice in \textit{ab initio} superconductivity predictions.

\begin{acknowledgments}
We thank Eva Kogler, Pedro N. Ferreira, Peter Cooke, Maélie Caussé, William Galloway, Stefano Racioppi and Johannes Meusburger for many helpful discussions. K.B. gratefully acknowledges support from the Engineering and Physical Sciences Research Council [Grant No. EP/W524633/1] through an EPSRC DTP studentship in the Department of Materials Science and Metallurgy. We acknowledge the EuroHPC Joint Undertaking for awarding this project access to the EuroHPC supercomputer LUMI, hosted by CSC (Finland) and the LUMI consortium through a EuroHPC Regular Access call.
\end{acknowledgments}

\section*{Data Availability}

The calculations in this work were performed with Mantel version 1.0~\cite{mantel-v1}. The latest version is available at \url{https://github.com/Kieran-Bozier/mantel}. The data that support the findings of this article are openly available~\cite{bozier2026data}.

\appendix


\bibliography{Mantel}

\end{document}